\documentclass[11pt,a4]{article}
\begin{document}

\title{{\bf Coherent and generalized intelligent states for infinite square well
potential and nonlinear oscillators}}
\author{A. H. El Kinani$^1$ and M. Daoud$^2$\vspace{0.5cm} \\
The Abdus Salam International Centre for Theoretical Physics, \\
ICTP-Strada Costiera 11, 34100 Trieste Italy\vspace{1cm}.\\
$^1$L.P.T, Physics Department, Faculty of Sciences, \\
University Mohammed V, P.O.Box 1014\\
Rabat, Morocco.\vspace{1cm}\\
$^2$L.P.M.C, Physics Department, Faculty of Sciences, \\
University Ibn Zohr, P.O.Box 28/S\\
Agadir, Morocco.\vspace{1.5cm}}
\date{}
\maketitle

\begin{abstract}
This article is an illustration of the construction of coherent and
generalized intelligent states which has been recently proposed by us for an
arbitrary quantum system $\left[ 1\right] $. We treat the quantum system
submitted to the infinite square well potential and the nonlinear
oscillators. By means of the analytical representation of the coherent
states \`{a} la Gazeau-Klauder and those \`{a} la Klauder-Perelomov, we
derive the generalized intelligent states in analytical ways.
\end{abstract}

\newpage

\section{Introduction}

The concept of coherent states (CS) has been successfully used in the last
decade in many different contexts of theoretical and experimental physics,
in particular quantum optics $\left[ 2-4\right] $. They were firstly
introduced for the harmonic oscillator (described by the Weyl-Heisenberg
algebra) by Schr\"{o}dinger $\left[ 5\right] .$

It is well known that, for the harmonic oscillator case, there are three
equivalent definitions of the coherent states $\left| z\right\rangle :\\{\bf %
D}_1:$ The elements of the set $\left\{ \left| z\right\rangle ,z\in {\bf C}%
\right\} $ are the eigenstates of the annihilation operator $a^{-}$%
\begin{equation}
a^{-}\left| z\right\rangle =z\left| z\right\rangle.
\end{equation}
${\bf D}_2:$ The coherent states $\left| z\right\rangle $ is the orbit of
the ground state $\left| 0\right\rangle $ under the Weyl-Heisenberg
displacement operator

\begin{equation}
\left| z\right\rangle =D\left( z\right) \left| 0\right\rangle
=\exp \left( za^{+}-\overline{z}a^{-}\right) \left|
0\right\rangle.
\end{equation}
Here $\left[ a^{-},a^{+}\right] ={\bf 1}$ and $\left( a^{-}\right) ^{\dagger
}=a^{+}.$ In view of this commutation relation, both the definitions ${\bf D}%
_1$ and ${\bf D}_2$ are equivalent.\\${\bf D}_3:$ Finally, the (CS) $\left|
z\right\rangle $ saturate the Heisenberg uncertainty relation

\begin{equation}
2\Delta X\Delta P=1
\end{equation}
with the position $X$ and momentum $P$ operators are given as usual by

\begin{equation}
X=\frac 1{\sqrt{2}}\left( a^{+}+a^{-}\right) \hbox{\hspace{1cm}and %
\hspace{1cm} }P=\frac i{\sqrt{2}}\left( a^{+}-a^{-}\right) .
\end{equation}
Furthermore, the (CS) $\left| z\right\rangle $ satisfy the following
properties: \\${\bf P}_1:$ The map $z\in {\bf C\rightarrow }\left|
z\right\rangle \in L^2\left( {\bf R}\right) $ is continuous. \\${\bf P}_2:$
The family of coherent states resolve the unity. Indeed, we have

\begin{equation}
\int\left| z\right\rangle \left\langle z\right| d\mu \left( z\right) =I,%
\hspace{1.5cm} d\mu \left( z\right) =\frac{d^2z}\pi =\frac 1\pi dRe%
\left( z\right) dIm \left( z\right).
\end{equation}
This property provide the useful analytic representation, known as the
Fock-Bargmann analytic representation, in which $a^{-}$ and $a^{+}$ are
represented respectively by $\partial _z$ and $z$ and the arbitrary state $%
\left| \psi \right\rangle $ is represented by the function $\psi \left(
z\right) =\exp \left( \frac{\left| z\right| ^2}2\right) \langle \overline{z}%
\left| \psi \right\rangle $ where $\overline{z}$ is the complex conjugate of
$z$.\\${\bf P}_3:$ The coherent states family is temporally stable. Indeed
\begin{equation}
e^{-iHt}\left| z\right\rangle =\left| ze^{-i\omega
t}\right\rangle.
\end{equation}
${\bf P}_4:$ They provide the classical-quantum correspondence.

The generalization of the above definitions for other potentials different
from the harmonic oscillator, was proposed recently by Gazeau and Klauder.
They give a general scheme leading to coherent states for an arbitrary
quantum system $\left[ 6\right] $ (see also $\left[ 7,8\right] $) by using
the definition ${\bf D}_1.$ A direct illustration of this construction was
given in $\left[ 9\right] $ for a particle trapped in the infinite square
well and in P\"{o}schl-Teller potentials. Using the definition ${\bf D}_2$,
the coherent states \`{a} la Klauder-Perelomov for the P\"{o}schl-Teller
potentials was given in $\left[ 10\right] $. To extend the third definition,
we have solved the eigenvalue equation of states minimizing the
Robertson-Schr\"{o}dinger uncertainty relation (which extend the Heisenberg
one) for an arbitrary quantum system $\left[ 1,11\right] $. The resulting
states are called the generalized intelligent states. We have shown that
these states includes the Gazeau-Klauder coherent one.

Recently, we gave the extension of the above three definitions for an
arbitrary quantum system and as an application, we treated a quantum system
evolving in the P\"{o}schl-Teller potentials $\left[ 12\right] $.

The main purpose of this paper is to give others illustrations for the
construction of coherent and generalized intelligent states for quantum
systems, trapped in the infinite square well potentials and nonlinear
oscillators.

We start by introducing in section 2, the Gazeau-Klauder coherent states
(definition ${\bf D}_1$), Klauder-Perelomov (definition ${\bf D}_2$) and
generalized intelligent states (definition ${\bf D}_3$) for an exact
solvable quantum system. Section 3 and 4 are respectively, devoted to the
infinite square well potential and the $x^4-$anharmonic oscillator.
Concluding remarks are given at the end of this work.

\section{General considerations}

In this section, we give the general scheme to follow in order to construct
the coherent states \`{a} la Gazeau-Klauder, \`{a} la Klauder-Perelomov and
the one's called generalized intelligent states for an arbitrary quantum
system.

\subsection{Gazeau-Klauder coherent states}

To begin, we choose a Hamiltonian $H$ admitting nondegenerate discrete
infinite spectrum $e_n$, such that the fundamental energy $e_0=0$ and the
others $\left\{ e_1,e_2....e_n\right\} $ are in increasing order i.e., $%
e_{n+1}>e_n$. The eigenstates $\left| \psi _n\right\rangle $ of $H$ are
orthogonal and satisfying

\begin{equation}
H\left| \psi _n\right\rangle =e_n\left| \psi _n\right\rangle .
\end{equation}
The Hamiltonian $H$ can be factorized as
\begin{equation}
H=A^{+}A^{-}
\end{equation}
where $A^{+}$ and $A^{-}$ are the creation and annihilation operators
respectively. They act on the states $\left| \psi _n\right\rangle $ as

\begin{eqnarray}
A^{-}\left| \psi _n\right\rangle &=&\sqrt{e_n}e^{i\alpha
(e_n-e_{n-1})}\left| \psi _{n-1}\right\rangle, \\ A^{+}\left| \psi
_n\right\rangle &=&\sqrt{e_{n+1}}e^{-i\alpha (e_{n+1}-e_n)}\left|
\psi _{n+1}\right\rangle ,
\end{eqnarray}
where $\alpha $ is a parameter belonging to ${\bf R.}$

The exponential factor appearing in the above equations produces only a
phase factor, and ensure, as we will see, the temporal stability of the
states which will be constructed in what follows.

The commutator of $A^{-}$ and $A^{+}$ takes the form

\begin{equation}
\left[ A^{-},A^{+}\right] =G(N)
\end{equation}
where $G(N)$ acts on $\left| \psi _n\right\rangle $ as
\begin{equation}
G(N)\left| \psi _n\right\rangle =(e_{n+1}-e_n)\left| \psi _n\right\rangle .
\end{equation}
Note that the operator $N\neq A^{+}A^{-}=H$ and it satisfies the following
commutation relations

\begin{equation}
\left[ A^{-},N\right] =A^{-},\hspace{0.7cm}\left[ A^{+},N\right]
=-A^{+}.
\end{equation}
The so-called Gazeau-Klauder coherent states $\left[ 6,7\right] $ are
defined as the eigenstates of the annihilation operators $A^{-}$. Let us
denote them by $\left| z,\alpha \right\rangle .$ They satisfy the eigenvalue
equation

\begin{equation}
A^{-}\left| z,\alpha \right\rangle =z\left| z,\alpha \right\rangle %
,\hspace{0.5cm}z\in {\bf C.}
\end{equation}
The solutions of $\left( 14\right) $ are given by

\begin{equation}
\left| z,\alpha \right\rangle ={\cal N(}\left| z\right| {\cal )}%
\sum\limits_{n=0}^{+\infty }\frac{z^ne^{-i\alpha
e_n}}{\sqrt{E(n)}}\left| \psi _n\right\rangle
\end{equation}
where the function $E(n)$ is defined by

\begin{equation}
E(n)=\left\{
\begin{array}{c}
1\hspace{2.3cm}\hbox{for }\hspace{0.5cm}n=0 \\
e_1e_2....e_n\hspace{1cm}\hbox{for }\hspace{0.5cm}n\neq 0
\end{array}
\right.
\end{equation}
and ${\cal N(}\left| z\right| {\cal )}$ the normalization
constant, which can be computed by using the normalization
condition $\left\langle z,\alpha \right. \left| \hbox{ }z,\alpha
\right\rangle =1.$ We obtain

\begin{equation}
{\cal N(}\left| z\right| {\cal )}=\left( \sum\limits_{n=0}^\infty \frac{%
\left| z\right| ^{2n}}{E\left( n\right) }\right) ^{-\frac 12}.
\end{equation}
It is then clear that the coherent states equation $\left( 15\right) $ are
continuous in $z\in {\bf C}$ and $\alpha \in {\bf R}$ and are temporally
stable under the evolution operator. Indeed

\begin{equation}
e^{-iHt}\left| z,\alpha \right\rangle =\left| z,\alpha +t\right\rangle .
\end{equation}
This property is ensured by the presence of the phase factor in equations $%
\left( 9\right) $ and $\left( 10\right) .$

In order to prove that the Gazeau-Klauder states resolves the identity, one
must find a measure $d\mu (z)$ such that
\begin{equation}
\int\left| z,\alpha \right\rangle \left\langle z,\alpha \right| \hbox{ }%
d\mu (z)=I_{{\cal H}}=\sum\limits_{n=0}^\infty \left| \psi
_n\right\rangle \left\langle \psi _n\right|
\end{equation}
where the integral is over the disk \{$z\in {\bf C,}\left| z\right| <{\cal R}
$\}, and ${\cal R}$ the radius of convergence defined as
\begin{equation}
{\cal R}\hbox{=}\lim\limits_{n\rightarrow \infty }\hbox{
}\sqrt[n]{E(n)}.
\end{equation}
Writing $d\mu (z)$\ as
\begin{equation}
d\mu (z)=\left[ {\cal N(}\left| z\right| {\cal )}\right] ^{-2}h(r^2)rdrd\phi %
,\hspace{0.7cm}z=re^{i\phi },
\end{equation}
and integrating over the whole plane, the resolution of the identity is then
equivalent to the determination of the function $h(u)$ satisfying
\begin{equation}
\int\nolimits_0^{+\infty }h(u)\hbox{ }u^{n-1}du=\frac{E(n-1)}\pi .
\end{equation}
So, it is clear that $h(u)$ is the inverse Mellin transform $\left[
13\right] $ of the function $\pi ^{-1}E(n-1)$

\begin{equation}
h(u)=\frac 1{2\pi i}\int\nolimits_{c-i\infty }^{c+i\infty }\frac{E(s-1)}%
\pi u^{-s}ds,\hspace{0.5cm}c\in {\bf R.}
\end{equation}
Then it is obvious that the explicit computation of the function
$h(u)$ requires the explicit knowledge of the spectrum of the
system under study. Two applications of such computation will be
given in sections 3 and 4.

Using Eq $\left( 14\right) $, one obtain the main value of the Hamiltonian $%
H $ in the states $\left| z,\alpha \right\rangle $%
\begin{equation}
\left\langle z,\alpha \right| H\left| z,\alpha \right\rangle
=\left| z\right| ^2.
\end{equation}
This relation is known as the action identity.

Finally, we remark that the coherent states $\left| z,\alpha \right\rangle $
can be written as an operator $U(z)$ acting on the ground state (up to
normalization constant) as
\begin{equation}
\left| z,\alpha \right\rangle =U(z)\left| \psi _0\right\rangle \hspace{1cm}%
\hbox{where }\hspace{0.5cm}U(z)=\exp \left( z\frac NHA^{+}\right)
.
\end{equation}
Note that the operator $U(z)$ is not unitary and can not be seen as a
displacement operator in the Klauder-Perelomov's sense. Hence, the resulting
coherent states can not be also interpreted as the Klauder-Perelomov's one.

\subsection{Klauder-Perelomov coherent states}

Following the definition ${\bf D}_2$, the coherent states of
Klauder-Perelomov type for an arbitrary quantum system are defined by
\begin{equation}
\left| z,\alpha \right\rangle =\exp
(zA^{+}-\overline{z}A^{-})\left| \psi _0\right\rangle
,\hspace{0.5cm}z\in {\bf C.}
\end{equation}
Using the action of the annihilation and creation operators on the Hilbert
space\\${\cal H}=\left\{ \left| \psi _n\right\rangle ,n=0,1,2,...\right\} $
one can show that the states $\left| z,\alpha \right\rangle $ can be written
as

\begin{equation}
\left| z,\alpha \right\rangle =\sum\limits_{n=0}^{+\infty
}z^nc_n\left( \left| z\right| \right) \sqrt{E(n)}e^{-i\alpha
e_n}\left| \psi _n\right\rangle .
\end{equation}
The quantities $c_n\left( \left| z\right| \right) $ are defined by
\begin{equation}
c_n\left( \left| z\right| \right) =\sum\limits_{j=0}^{+\infty
}\frac{\left( -\left| z\right| ^2\right) ^j}{\left( n+2j\right)
!}\pi (n+1,j)
\end{equation}
where
\begin{equation}
\pi
(n+1,j)=\sum\limits_{i_1=1}^{n+1}e_{i_1}\sum%
\limits_{i_2=1}^{i_1+1}e_{i_2}....\sum\limits_{i_j=1}^{i_{j-1}+1}e_{i_j}%
,\hspace{1cm}\pi \left( n+1,0\right) =1.
\end{equation}
One can verify that the $\pi $'s satisfy the following relation
\begin{equation}
\pi \left( n+1,j\right) -\pi \left( n,j\right) =e_{n+1}\pi \left(
n+2,j-1\right) .
\end{equation}
Using this recurrence formula, it is not difficult to show that the $%
c_n\left( \left| z\right| =r\right) ^{\prime }$s satisfy the following
differential equation
\begin{equation}
r\frac{dc_n\left( r\right) }{dr}=c_{n-1}\left( r\right) -nc_n\left( r\right)
-e_{n+1}r^2c_{n+1}\left( r\right) .
\end{equation}
The solution of this differential equation lead to the explicit expression
of the coherent states of Klauder-Perelomov's type for an arbitrary quantum
system. Here also, we note that the solutions are intimately related to the
spectrum of the system under study. Solutions in the cases corresponding to
nonlinear oscillators (with an $x^4-$interaction) and the infinite square
will potential will be treated in the sequel of this paper.

\subsection{Generalized intelligent states}

These states are known as those minimizing the so-called
Robertson-Schr\"{o}dinger uncertainty relation $\left[ 14,15\right] $ (for
more details see $\left[ 1\right] $ where they were constructed for an exact
solvable system) and generalize the Gazeau-Klauder ones. In what follows, we
give a short review of there main characteristics.

We introduce two hermitian operators defined in terms of the operators $%
A^{-} $ and $A^{+}$ as follows

\begin{equation}
W=\frac 1{\sqrt{2}}(A^{-}+A^{+}),\hspace{0.5cm}P=\frac i{\sqrt{%
2}}(A^{+}-A^{-})
\end{equation}
which satisfy the commutation relation

\begin{equation}
\left[ W,P\right] =iG(N)\equiv iG
\end{equation}
where the operator $G$ is defined by $\left( 12\right) .$

It is well known that for two hermitian operators $W$ and $P$ satisfying the
non-canonical commutation relation $\left( 33\right) $, the variances $%
\left( \Delta W\right) ^2$ and $\left( \Delta P\right) ^2$ satisfy the
Robertson-Schr\"odinger uncertainty relation

\begin{equation}
\left( \Delta W\right) ^2\left( \Delta P\right) ^2\geq \frac 14\left(
\left\langle G\right\rangle ^2+\left\langle F\right\rangle ^2\right)
\end{equation}
where the operator $F$ is defined by

\begin{equation}
F=\left\{ W-\left\langle W\right\rangle ,P-\left\langle P\right\rangle
\right\}
\end{equation}
and its mean value is expressed by

\begin{equation}
\left\langle F\right\rangle =i\left[ \left( \Delta A^{+}\right) ^2-\left(
\Delta A^{-}\right) ^2\right]
\end{equation}
in terms of the variances of $A^{-}$ and $A^{+}$.

The symbol $\left\{ ,\right\} ,$ appearing in equation $\left( 35\right) ,$
stands for the anticommutator and the uncertainty relation $\left( 34\right)
$ is a generalization of the well known Heisenberg one.

The so-called generalized intelligent states are obtained when the equality
in the Robertson-Schr\"odinger uncertainty relation is realized $\left[
16\right] $ (see also $\left[ 17,18\right] $)$.$ They satisfy the eigenvalue
equations
\begin{equation}
\left( W+i\lambda P\right) \left| z,\lambda ,\alpha \right\rangle =z\sqrt{2}%
\left| z,\lambda ,\alpha \right\rangle,\hspace{0.5cm}\lambda ,z\in
{\bf C.}
\end{equation}
Using the equation $\left( 32\right) $, the above equation can be rewritten
as
\begin{equation}
\left[ \left( 1-\lambda \right) A^{+}+\left( 1+\lambda \right) A^{-}\right]
\left| z,\lambda ,\alpha \right\rangle =2z\left| z,\lambda ,\alpha
\right\rangle .
\end{equation}
For the generalized intelligent states, solutions of $\left( 38\right) $,
the variances of $W$ and $P$ are

\begin{equation}
\left( \Delta W\right) ^2=\left| \lambda \right| \Delta,%
\hspace{0.6cm}\left( \Delta P\right) ^2=\frac 1{\left| \lambda \right|
}\Delta
\end{equation}
with
\begin{equation}
\Delta =\frac 12\sqrt{\left\langle G\right\rangle ^2+\left\langle
F\right\rangle ^2}.
\end{equation}
The main values of $G$ and $F$, in the generalized intelligent states can be
expressed in terms of the variances of $P$ as follows
\begin{eqnarray}
\left\langle G\right\rangle &=&2Re(\lambda )\left( \Delta P\right)
^2,
\\
\left\langle F\right\rangle &=&2Im(\lambda )\left( \Delta P\right)
^2.
\end{eqnarray}
Clearly, for $\left| \lambda \right| =1$, we have

\begin{equation}
\left( \Delta W\right) ^2=\left( \Delta P\right) ^2.
\end{equation}
The states satisfying $\left( 38\right) ,$ with $\left| \lambda \right| =1,$
are called the generalized coherent states. The complete classification of
the solutions of $\left( 38\right) $ was considered in $\left[ 1\right] $
for an exact solvable quantum system. In what follows we give the main
results of this classification which will be adopted to the quantum systems
considered in this work.

Solutions of the equation $\left( 38\right) $ for $\lambda \neq -1,$ are
given by
\begin{equation}
\left| z,\lambda ,\alpha \right\rangle =\sum\limits_{n=0}^{+\infty
}a_n(z)\left| \psi _n\right\rangle
\end{equation}
where
\begin{equation}
a_n(z)=a_0\frac{\left( 2z\right) ^n}{\left( 1+\lambda \right) ^n\sqrt{E(n)}}%
\left[ \sum\limits_{h=0\left( 1\right) \left[ \frac n2\right]
}\left(
-1\right) ^h\frac{\left( 1-\lambda ^2\right) ^h}{\left( 2z\right) ^{2h}}%
\Delta \left( n,h\right) \right] e^{-i\alpha e_n}.
\end{equation}
The symbol $\left[ \frac n2\right] $ stands for the integer part of $\frac
n2 $ and the function $\Delta \left( n,h\right) $ is defined by
\begin{equation}
\Delta \left( n,h\right) =\sum\limits_{j_1=1}^{n-\left(
2h-1\right) }e_{j_1}\left[ \sum\limits_{j_2=j_1+2}^{n-\left(
2h-3\right) }e_{j_2}...\left[ ...\left[
\sum\limits_{j_h=j_{h-1}+2}^{n-1}e_{j_h}\right] ...\right]
...\right] .
\end{equation}
The states $\left| z,\lambda ,\alpha \right\rangle $ can be written, in a
compact form, as the action of the operator $U(\lambda ,z)$ on the ground
state $\left| \psi _0\right\rangle $ of $H$ as

\begin{equation}
\left| z,\lambda ,\alpha \right\rangle =U(\lambda ,z)\left| \psi
_0\right\rangle
\end{equation}
where $U(\lambda ,z)$ is defined (up to normalization constant) as

\begin{equation}
U\left( z,\lambda \right) =\sum_{n=0}^{+\infty }\left[ \left( \frac{2z}{%
1+\lambda }\right) \frac 1HA^{+}+\left( \frac{\lambda -1}{\lambda
+1}\right) \frac 1H\left( A^{+}\right) ^2\right] ^n.
\end{equation}
For more details, we invite the reader to see the reference $\left[ 1\right]
$. It is clear that the generalized intelligent states $\left( 44\right) $
obtained by the minimization of the Robertson-Schr\"odinger uncertainty
relation are different from the others introduced before, for an arbitrary
quantum system, the definitions ${\bf D}_1,$ ${\bf D}_2$ and ${\bf D}_3$
leads to inequivalent families of states except, of course, for the harmonic
oscillator case. All these matters will be adapted in what follows to two
interesting quantum mechanical systems: the infinite square well potential
and the anharmonic oscillators.

\section{Infinite square well potential}

In this section, we will give the coherent and generalized intelligent
states for a quantum system trapped in an infinite square well potential by
exploiting the results of the previous section.

\subsection{Gazeau-Klauder coherent states}

Let us recall the eigenvalues and eigenvectors of the Hamiltonian $H$
corresponding to a quantum system submitted to the infinite square well
potential. Indeed, $H$ acts on $\left| \psi _n\right\rangle $ as

\begin{equation}
H\left| \psi _n\right\rangle =e_n\left| \psi _n\right\rangle \hspace{0.5cm}%
\hbox{where}\hspace{0.5cm}e_n=n(n+2).
\end{equation}
The lowering and raising operators $A^{-}$ and $A^{+}$act on $\left| \psi
_n\right\rangle $ now as follows

\begin{eqnarray}
A^{-}\left| \psi _n\right\rangle &=&\sqrt{n(n+2)}e^{i\alpha
(2n+1)}\left| \psi _{n-1}\right\rangle, \\ A^{+}\left| \psi
_n\right\rangle &=&\sqrt{(n+1)(n+3)}e^{-i\alpha (2n+3)}\left| \psi
_{n+1}\right\rangle
\end{eqnarray}
and the Hamiltonian $H$ can be factorized as

\begin{equation}
H=A^{+}A^{-}.
\end{equation}
The number operator ($N\neq A^{+}A^{-}=H$) acts on $\left| \psi
_n\right\rangle $ as follows

\begin{equation}
N\left| \psi _n\right\rangle =n\left| \psi _n\right\rangle
\end{equation}
and the commutation relation between $A^{-}$ and $A^{+}$ is given by

\begin{equation}
\left[ A^{-},A^{+}\right] =G(N)
\end{equation}
where $G(N)$ is defined as

\begin{equation}
G(N)=\left( 2N+3\right).
\end{equation}
The Hilbert space ${\cal H}$ for the infinite square well potential is
easily constructed in the same way as the standard harmonic oscillator. This
space is spanned by the states

\begin{equation}
\left| \psi _n\right\rangle =\frac{\left( A^{+}\right) ^n}{\sqrt{E(n)}}%
e^{i\alpha e_n}\left| \psi _0\right\rangle
\end{equation}
where
\begin{equation}
E(n)=\frac{n!\left( n+2\right) !}2.
\end{equation}
The Gazeau-Klauder coherent states equation $\left( 15\right) $ becomes
\begin{equation}
\left| z,\alpha \right\rangle ={\cal N(}\left| z\right| {\cal )}%
\sum\limits_{n=0}^{+\infty }\frac{z^n\sqrt{2}e^{-i\alpha n(n+2)}}{\sqrt{%
\Gamma \left( n+1\right) \Gamma \left( n+3\right) }}\left| \psi
_n\right\rangle
\end{equation}
where the normalization constant is

\begin{equation}
{\cal N(}\left| z\right| {\cal )}=\left[ _0F_1\left( 3,\left| z\right|
^2\right) \right] ^{-\frac 12}.
\end{equation}
The Gazeau-Klauder coherent states for the system under study are normalized
but they are not orthogonal to each other. Indeed, we have
\begin{equation}
\left\langle z,\alpha \right| z^{\prime },\alpha \rangle =\frac{_0F_1(3,%
\overline{z}z^{\prime })}{\sqrt{_0F_1\left( 3,\left| z\right|
^2\right) \hbox{ }_0F_1\left( 3,\left| z^{\prime }\right|
^2\right) }}
\end{equation}
The set of states $\left( 58\right) $ are overcomplet in respect to the
measure

\begin{equation}
d\mu (z)=\frac 2\pi I_2(2r)K_1(2r)rdrd\phi,\hspace{0.7cm}%
z=re^{i\phi }
\end{equation}
where $I_\nu (x)$ and $K_\nu (x)$ are respectively the modified Bessel
functions of the first and second kinds.

By using this last property, one can represent the state space as
the Hilbert space of analytic function in the whole plane. So, for
a normalized state \\$\left| \Psi \right\rangle
=\sum\limits_{n=0}^{+\infty }b_n\left| \psi _n\right\rangle $ one
gets

\begin{eqnarray}
\Psi (z,\alpha ) &\equiv &\sqrt{_0F_1(3,\left| z\right| ^2)}\langle
\overline{z},\alpha \left| \Psi \right\rangle  \nonumber \\
&=&\sum\limits_{n=0}^{+\infty }b_n\frac{z^n\sqrt{2}e^{i\alpha n(n+2)}}{%
\sqrt{\Gamma \left( n+1\right) \Gamma \left( n+3\right) }}.
\end{eqnarray}
Then, it is obvious that for the state $\left| \psi _n\right\rangle $ we
associate

\begin{equation}
\psi _n(z,\alpha )=\frac{z^n\sqrt{2}e^{i\alpha
n(n+2)}}{\sqrt{\Gamma \left( n+1\right) \Gamma \left( n+3\right)
}}.
\end{equation}
The operators $A^{-},$ $A^{+}$ and $G(N)$ act on the Hilbert space of
analytic functions as first order differential operators

\begin{equation}
A^{+}=z,\hspace{0.7cm}A^{-}=z\frac{d^2}{dz^2}+3\frac d{dz}%
\hspace{0.5cm}\hbox{and}\hspace{0.5cm}G(N)=2z\frac d{dz}+3
\end{equation}
It is easy to verify that the actions of $A^{+},$ $A^{-}$ and $G\left(
N\right) $ on $\psi _n(z,\alpha )$ lead to

\begin{eqnarray}
A^{+}\psi _n(z,\alpha ) &=&\sqrt{\left( n+1\right) \left( n+3\right) }%
e^{-i\alpha (2n+3)}\psi _{n+1}\left( z,\alpha \right), \\
A^{-}\psi _n(z,\alpha ) &=&\sqrt{n\left( n+2\right) }e^{i\alpha
(2n+1)}\psi _{n-1}\left( z,\alpha \right), \\ G\left( N\right)
\psi _n(z,\alpha ) &=&(2n+3)\psi _n\left( z,\alpha \right) .
\end{eqnarray}
It is obvious that the coherent states constructed here are
temporally stable and satisfying Eq. $\left( 24\right) $.

Finally, we remark that the coherent states $\left| z,\alpha \right\rangle $
can be written as an operator $U(z)$ acting on the ground state $\left| \psi
_0\right\rangle $ (up to normalization constant)
\begin{equation}
U(z)=\exp \left( z\frac 1{N+2}A^{+}\right)
\end{equation}
such that we have
\begin{equation}
\left| z,\alpha \right\rangle =U(z)\left| \psi _0\right\rangle .
\end{equation}
As we mentioned, in the section 1, the operator $U(z)$ Eq. $\left( 68\right)
$ is not unitary. The analytic representation of the coherent states \`a la
Gazeau-Klauder introduced here are important since it will be used to derive
the generalized intelligent states in an analytical way.

\subsection{Klauder-Perelomov coherent states}

The coherent states \`{a} la Klauder-Perelomov far an arbitrary quantum
system are defined in the subsection $\left( 2.2\right) .$ We have shown
that their explicit expressions depends on the spectrum structure of the
system. Here, we will solve the differential equation $\left( 31\right) $
for the spectrum of the infinite square well potential. Indeed for $%
e_n=n(n+2)$ the $c_n$ 's coefficient admit the solution

\begin{equation}
c_n(r)=\frac 1{n!}\left( \cosh (r)\right) ^{-3}\left( \frac{\tanh \left(
r\right) }r\right) ^n.
\end{equation}
The coherent states of Klauder-Perelomov's type takes the form
\begin{equation}
\left| z,\alpha \right\rangle =\left( 1-\tanh ^2\left( \left|
z\right| \right) \right) ^{\frac 32}\sum_{n=0}^{+\infty }\left(
\frac{z\tanh \left( \left| z\right| \right) }{\left| z\right|
}\right) ^n\left[ \frac{\left( n+1\right) \left( n+2\right)
}2\right] ^{\frac 12}e^{-i\alpha n(n+2)}\left| \psi
_n\right\rangle .
\end{equation}
By setting $\zeta =\frac{z\tanh \left( \left| z\right| \right) }{\left|
z\right| }$, we obtain
\begin{equation}
\left| \zeta ,\alpha \right\rangle \equiv \left( 1-\left| \zeta \right|
^2\right) ^{\frac 32}\sum\limits_{n=0}^{+\infty }\zeta ^n\left[ \frac{%
\left( n+1\right) \left( n+2\right) }2\right] ^{\frac 12}e^{-i\alpha
n(n+2)}\left| \psi _n\right\rangle .
\end{equation}
We note that the parameter $\zeta $ belongs to the unit disk
$D=\left\{ \zeta {\bf \in C},\hbox{ }\left| \zeta \right|
<1\right\} .$

The states $\left| \zeta ,\alpha \right\rangle ,$ are temporally stable.
Indeed we have
\begin{equation}
e^{-iHt}\left| \zeta ,\alpha \right\rangle =\left| \zeta ,\alpha
+t\right\rangle .
\end{equation}
From the Eq. $\left( 72\right) ,$ one can see that the
Klauder-Perelomov coherent states are normalized but not
orthogonal to each others
\begin{equation}
\left\langle \zeta ,\alpha \right| \zeta ^{^{\prime }},\alpha ^{^{\prime
}}\rangle =\sqrt{\left( 1-\left| \zeta \right| ^2\right) ^3\left( 1-\left|
\zeta ^{^{\prime }}\right| ^2\right) ^3}\sum\limits_{n=0}^{+\infty }\frac{%
\left( \overline{\zeta }\zeta ^{^{\prime }}\right) ^n}{n!}\frac{\Gamma
\left( n+3\right) }2e^{-i(\alpha ^{^{\prime }}-\alpha )n(n+2)}.
\end{equation}
The measure ensuring the identity resolution of $\left| \zeta ,\alpha
\right\rangle $ takes the form
\begin{equation}
d\mu \left( \zeta \right) =\frac 2\pi \frac{d^2\zeta }{\left( 1-\left| \zeta
\right| ^2\right) ^2}.
\end{equation}
Then we can express any coherent states in terms of the others
\begin{equation}
\mid \zeta ^{^{\prime }},\alpha ^{^{\prime }}\rangle =\int\left|
\zeta ,\alpha \right\rangle \left\langle \zeta ,\alpha \right|
\zeta ^{^{\prime }},\alpha ^{^{\prime }}\rangle d\mu \left( \zeta
\right) .
\end{equation}
For any state $\left| \Phi \right\rangle
=\sum\limits_{n=0}^{+\infty }c_n\left| \psi _n\right\rangle $ in
the Hilbert space, one can construct the analytic function:

\begin{equation}
\Phi (\zeta {\bf ,}\alpha )=\left( 1-\left| \zeta \right|
^2\right) ^{-\frac 32}\left\langle \overline{\zeta },\alpha
\right| \Psi \rangle =\sum\limits_{n=0}^{+\infty }\zeta
^n\sqrt{\frac{(n+1)(n+2)}2}c_ne^{i\alpha n(n+2)}.
\end{equation}
Here, the $\left| \psi _n\right\rangle $ state is represented by the function

\begin{equation}
\psi _n^{\prime }(\zeta ,\alpha )=\zeta ^n\sqrt{\frac{(n+1)(n+2)}2}%
e^{i\alpha n(n+2)}.
\end{equation}
The operators $A^{\pm }$ and $G(N)$ act on the Hilbert space of analytic
functions $\Phi (\zeta {\bf ,}\alpha )$ as first-order differential operators

\begin{equation}
A^{+}=\zeta ^2\frac d{d\zeta }+3\zeta,\hspace{0.5cm}%
A^{-}=\frac d{d\zeta
}\hspace{0.5cm}\hbox{and}\hspace{0.5cm}G(N)=2\zeta \frac d{d\zeta
}+3
\end{equation}
By a simple computation, one can verify

\begin{eqnarray}
A^{+}\psi _n^{\prime }(\zeta ,\alpha ) &=&\sqrt{\left( n+1\right)
\left( n+3\right) }e^{-i\alpha (2n+3)}\psi _{n+1}^{\prime }(\zeta
,\alpha ), \\
A^{-}\psi _n^{\prime }(\zeta ,\alpha ) &=&\sqrt{n\left( n+2\right) }%
e^{i\alpha (2n+1)}\psi _{n-1}^{\prime }(\zeta ,\alpha ), \\
G\left( N\right) \psi _n^{\prime }(\zeta ,\alpha ) &=&(2n+3)\psi
_n^{\prime }(\zeta ,\alpha ).
\end{eqnarray}
Finally, we note that the above analytic representations will be the main
tool by means of which we can get the analytic solutions of generalized
intelligent states.

\subsection{Generalized intelligent states}

In this part, we will use the analytic representation of the coherent states
introduced before in the subsections $\left( 3.1\right) $ and $\left(
3.2\right) $, in order to obtain the generalized intelligent states in an
analytical way.

\subsubsection{Gazeau-Klauder analytic representation}

In this representation, we define the Hilbert space as a space of functions $%
S$ which are holomorphic in the complex plane. The scalar product is given by

\begin{equation}
\langle f\left| g\right\rangle =\int\overline{f(z)}g(z)d\mu (z)
\end{equation}
where $d\mu (z)$ is the measure defined by Eq. $\left( 61\right)
.$

By introducing the analytic function
\begin{equation}
\Psi _{(z^{\prime },\lambda ,\alpha )}(z)=\sqrt{_0F_1(3,\left| z\right| ^2)}%
\langle \overline{z},\alpha \left| z^{\prime },\lambda ,\alpha \right\rangle
\end{equation}
we can convert the eigenvalue equation

\begin{equation}
\left[ \left( 1-\lambda \right) A^{+}+\left( 1+\lambda \right) A^{-}\right]
\left| z^{\prime },\lambda ,\alpha \right\rangle =2z^{\prime }\left|
z^{\prime },\lambda ,\alpha \right\rangle
\end{equation}
into the second-order linear homogeneous differential equation
\begin{equation}
\left[ \left( 1+\lambda \right) \left( z\frac{d^2}{dz^2}+3\frac d{dz}\right)
+\left( 1-\lambda \right) z-2z^{\prime }\right] \Psi _{(z^{\prime },\lambda
)}\left( z\right) =0.
\end{equation}
Firstly, we consider the general case $\lambda \neq \pm 1$. Setting

\begin{equation}
\Psi _{(z^{\prime },\lambda )}\left( z\right) =\exp \left( \pm \sqrt{\frac{%
\lambda -1}{\lambda +1}}z\right) F_{(z^{\prime },\lambda )}(z),
\end{equation}
one can transformed  $\left( 86\right) $ into the Kummer equation
\begin{equation}
\left[ Z\frac{d^2}{dZ^2}+\left( 3-Z\right) \frac d{dZ}-\left(
\frac 32\mp \frac{z^{\prime }}{\sqrt{\lambda ^2-1}}\right) \right]
F_{(z^{\prime },\lambda )}\left( z\right) =0
\end{equation}
where $Z=\mp 2\sqrt{\frac{\lambda -1}{\lambda +1}}z.$

Then the solutions of the equation $\left( 86\right) $ are given by
\begin{equation}
\Psi _{(z^{\prime },\lambda )}\left( z\right) =\exp \left( \pm \sqrt{\frac{%
\lambda -1}{\lambda +1}}z\right) \hbox{ }_1F_1\left( \frac 32\mp \frac{%
z^{\prime }}{\sqrt{\lambda ^2-1}},3;\mp 2\sqrt{\frac{\lambda -1}{\lambda +1}%
}z\right)
\end{equation}
or
\begin{equation}
\Psi _{(z^{\prime },\lambda )}\left( z\right) =\exp \left( \pm \sqrt{\frac{%
\lambda -1}{\lambda +1}}z\right) z^{-2}\hbox{ }_1F_1\left( -\frac
12\mp
\frac{z^{\prime }}{\sqrt{\lambda ^2-1}},-1;\mp 2\sqrt{\frac{\lambda -1}{%
\lambda +1}}z\right) .
\end{equation}
The first solution $\left( 89\right) $ is always analytic, but the second $%
\left( 90\right) $ is not. Since the hypergeometric function
$_1F_1(a;b;z)$ satisfies the equation
\begin{equation}
_1F_1(a;b;z)=e^z\hbox{ }_1F_1(b-a;b;-z),
\end{equation}
the upper and lower signs in equation $\left( 89\right) $ are equivalent.

By using the properties of this hypergeometric functions, we conclude that
the squeezing parameter $\lambda $ obeys to the condition

\begin{equation}
\sqrt{\left| \frac{\lambda -1}{\lambda +1}\right| }<1\Leftrightarrow
Re(\lambda )>0
\end{equation}
which traduce the restriction on $\lambda $ imposed by the positivity of the
commutator $\left[ A^{-},A^{+}\right] =2N+3$ (see equations $\left(
53\right) $ and $\left( 54\right) $).

We consider now the degenerate cases $\lambda =\pm 1.$ For the case $\lambda
=-1$ the equation $\left( 86\right) $ does not have any normalized analytic
solution (the operator $A^{+}$ does not have any eigenstate). For $\lambda
=1,$ using the power series of $_1F_1(a,b;z)$, we get
\begin{equation}
\Psi _{(z^{\prime },\lambda =1)}\left( z\right) =\hbox{
}_0F_1(3;zz^{\prime }).
\end{equation}
The result $\left( 93\right) $ coincides with the solution $\left( 58\right)
$ (up to normalization constant). Then we recover the infinite square well
coherent states defined as the $A^{-}$ eigenstates (Gazeau-Klauder coherent
states).

\subsubsection{Klauder-Perelomov analytic representation}

In this representation the Hilbert space is equipped with the following
scalar product
\begin{equation}
\langle f\left| g\right\rangle =\int\overline{f(\zeta )}g(\zeta
)d\mu (\zeta ).
\end{equation}
Note that the integration is over the unit disk $D=\left\{ \zeta {\bf \in C},%
\hbox{ }\left| \zeta \right| <1\right\} $ and the measure is
defined by equation $\left( 75\right) .$

To solve the eigenvalue equation $\left( 38\right) $ we introduce the
analytic function
\begin{equation}
\Phi _{(\zeta ^{\prime },\lambda )}\left( \zeta \right) =\sqrt{\left(
1-\left| \zeta \right| ^2\right) ^{-3}}\langle \overline{\zeta },\alpha
\left| \zeta ^{\prime },\lambda ,\alpha \right\rangle .
\end{equation}
Then $\left( 38\right) $ is converted to the following
differential equation

\begin{equation}
\left[ \left[ (1-\lambda )\zeta ^2+(1+\lambda )\right] \frac d{d\zeta
}+3(1-\lambda )\zeta -2\zeta ^{\prime }\right] \Phi _{(\zeta ^{\prime
},\lambda )}(\zeta )=0.
\end{equation}
Admissible values of $\lambda $ and $\zeta ^{\prime }$ are
determined by the requirements that the functions $\Phi _{(\zeta
^{\prime },\lambda )}(\zeta )$ should be analytic in the unit
disk. The solutions of the Eq. $\left( 96\right) $ are

\begin{equation}
\Phi _{(\zeta ^{\prime },\lambda )}(\zeta )={\cal A}\left( \left| \zeta
\right| \right) \left( 1+\left( \frac{\lambda -1}{\lambda +1}\right) ^{\frac
12}\zeta \right) ^{\alpha _{+}}\left( 1-\left( \frac{\lambda -1}{\lambda +1}%
\right) ^{\frac 12}\zeta \right) ^{\alpha _{-}}
\end{equation}
where
\begin{equation}
\alpha _{\pm }=-\frac 32\pm \frac{\zeta ^{\prime }}{\sqrt{\lambda ^2-1}}
\end{equation}
and ${\cal A}\left( \left| \zeta \right| \right) $ is a normalization
constant. The condition of analyticity requires

\begin{equation}
\left| \frac{\lambda -1}{\lambda +1}\right| <1\Leftrightarrow
\hbox{ }Re\lambda >0.
\end{equation}
If Re$\lambda <0$, the functions $\Phi _{(\zeta ^{\prime },\lambda )}(\zeta
) $ cannot be analytic in the unit disk.

The decomposition of the generalized intelligent states $\left| \zeta
^{\prime },\lambda ,\alpha \right\rangle $ over the Hilbert orthonormal
basis \{$\left| \psi _n\right\rangle $\} can be obtained by expanding the
function $\Phi _{(\zeta ^{\prime },\alpha )}(\zeta )$ into a power series in
$\zeta .$ This can be done by using the following relations

\begin{equation}
\prod\limits_{l=\pm 1}\left( 1+\left( \frac{\lambda -1}{\lambda
+1}\right) ^{\frac 12}\zeta \right) ^{-\frac 32+l\frac{\zeta
^{\prime }}{\sqrt{\lambda
^2-1}}}=\sum\limits_{n=0}^{+\infty }\zeta ^n\left( 2\sqrt{\frac{\lambda -1}{%
\lambda +1}}\right) ^nP_n^{(\alpha _{+}-n,\alpha _{-}-n)}(0).
\end{equation}
Then, the function $\Phi _{(\zeta ^{^{\prime }},\alpha )}(\zeta )$ can be
expanded in terms of the Jacobi polynomials $P_n^{(\alpha ,\beta )}(x)$.
Using the relation between the hypergeometric function and Jacobi
polynomials $\left[ 19\right] $ we can show that

\begin{eqnarray}
\left| \zeta ^{\prime },\lambda ,\alpha \right\rangle &=&{\cal A}\left(
\left| \zeta \right| \right) \sum\limits_{n=0}^{+\infty }\left[ \frac{n!}{%
(n+2)!}\right] ^{\frac 12}\left[ \frac{n!\Gamma (\alpha _{+}-n+1)}{\Gamma
(\alpha _{+}+1)}\right] \left( 2\sqrt{\frac{\lambda -1}{\lambda +1}}\right)
^n\times  \nonumber \\
&&_2F_1(-n,-n-2;\alpha _{+}-n+1;\frac 12)e^{-i\alpha n(n+2)}\left| \psi
_n\right\rangle
\end{eqnarray}
or
\begin{equation}
\left| \zeta ^{\prime },\lambda ,\alpha \right\rangle ={\cal A}\left( \left|
\zeta \right| \right) \sum\limits_{n=0}^{+\infty }\left[ \frac{n!}{(n+2)!}%
\right] ^{\frac 12}\left( 2\sqrt{\frac{\lambda -1}{\lambda +1}}\right)
^nP_n^{(\alpha _{+}-n,\alpha _{-}-n)}(0)e^{-i\alpha n(n+2)}\left| \psi
_n\right\rangle .
\end{equation}
The generalized intelligent states $\Phi _{(\zeta ^{\prime
},\lambda )}(\zeta {\bf )}$ (Eq. $\left( 97\right) $) and $\Psi
_{(z^{\prime },\lambda )}(z{\bf )}$ (Eq. $\left( 89\right) $) are
related through a Laplace transform $\left[ 20\right] $. In fact,
Eq. $\left( 96\right) $ can be written as

\begin{equation}
\left[ \left[ (1+\lambda )\zeta ^2+(1-\lambda )\right] \frac d{d\zeta }-3%
\frac{(1-\lambda )}\zeta +2\zeta ^{\prime }\right] \Phi _{(\zeta
^{\prime },\lambda )}\left( \frac 1\zeta \right) =0.
\end{equation}
Using
\begin{equation}
\Phi _{(\zeta ^{\prime },\lambda )}\left( \frac 1\zeta \right)
=\frac{\zeta ^{-3}}{\sqrt{2}}\int\nolimits_0^{+\infty }z^2\Psi
_{(\zeta ^{\prime },\lambda )}(z{\bf )}e^{-\frac z\zeta }dz,
\end{equation}
it is easy to see that the eigenvalue equation $\left( 103\right) $ becomes

\begin{equation}
\left[ \left( 1+\lambda \right) \left( z\frac{d^2}{dz^2}+3\frac d{dz}\right)
+\left( 1-\lambda \right) z-2\zeta ^{\prime }\right] \Psi _{(\zeta ^{\prime
},\lambda )}\left( z\right) =0.
\end{equation}
We note that this last differential equation coincides with
$\left( 86\right) $ for $\zeta ^{\prime }=z^{\prime }$.

\section{$x^4-$anharmonic oscillator}

Let us recall briefly the general structure of the Hamiltonian eigenvalues
and eigenvectors for the one-dimensional nonlinear oscillators. Indeed, we
are interested to the Hamiltonian which has the form

\begin{equation}
H=a^{+}a^{-}+\frac \varepsilon 4\left( a^{-}+a^{+}\right) ^4-c_0
\end{equation}
where $a^{+}$ and $a^{-}$ are the creation and annihilation operators for
the harmonic oscillator and the parameter $\varepsilon $ is positive. The
quantity $c_0$ is defined as follows

\begin{equation}
c_0=\frac 34\varepsilon -\frac{21}8\varepsilon ^2.
\end{equation}
The Hamiltonian $H$ can be factorized in the following form $\left[
21\right] $

\begin{equation}
H=A_\varepsilon ^{+}A_\varepsilon ^{-}
\end{equation}
where $(A_\varepsilon ^{-})^{\dagger }=A_\varepsilon ^{+}$ and the operator $%
A_\varepsilon ^{-}$ is defined as a some function of $a^{-}$ and
$a^{+}$ (for more detail see $\left[ 21\right]) .$

The energy levels are given by $\left[ 21\right] $ (see also $\left[
22\right] $)

\begin{equation}
e_n=n+\frac 32\varepsilon \left( n^2+n\right) .
\end{equation}
The Hilbert space ${\cal H}$ is spanned by the states

\begin{equation}
\left| n,\varepsilon \right\rangle =\frac{\left( A_\varepsilon ^{+}\right) ^n%
}{\sqrt{E(n)}}e^{i\alpha e_n}\left| 0,\varepsilon \right\rangle %
,\hspace{1cm}n\in {\bf N}
\end{equation}
where $\left| 0,\varepsilon \right\rangle $ is the ground state and the
function $E(n)$ is defined by

\begin{equation}
E(n)=\left\{
\begin{array}{c}
\hspace{0.5cm}1\hbox{\hspace{3.5cm}if }\hspace{1cm}n=0 \\ \left(
\frac{3\varepsilon }2\right) ^n\frac{\Gamma (n+1)\Gamma (n+2+\frac
2{3\varepsilon })}{\Gamma (2+\frac 2{3\varepsilon })}\hbox{\hspace{0,6cm}if %
\hspace{1cm}}n\neq 0.
\end{array}
\right.
\end{equation}
The action of the annihilation and creation operators are defined as follows:

\begin{eqnarray}
A_\varepsilon ^{+}\left| n,\varepsilon \right\rangle &=&\sqrt{\left( \frac{%
3\varepsilon }2\right) \left( n+1\right) \left( n+2+\frac 2{3\varepsilon
}\right) }e^{-i\alpha (e_{n+1}-e_n)}\left| n+1,\varepsilon \right\rangle \\
A_\varepsilon ^{-}\left| n,\varepsilon \right\rangle &=&\sqrt{\left( \frac{%
3\varepsilon }2\right) n\left( n+1+\frac 2{3\varepsilon }\right) }e^{i\alpha
(e_n-e_{n-1})}\left| n-1,\varepsilon \right\rangle .
\end{eqnarray}
We define the number operator $N$ as
\begin{equation}
N\left| n,\varepsilon \right\rangle =n\left| n,\varepsilon \right\rangle
\end{equation}
The operator $N$ is different from the product $A_\varepsilon
^{+}A_\varepsilon ^{-}(=H)$.

\subsection{Gazeau-Klauder coherent states}

Following the construction introduced before, the Gazeau-Klauder coherent
states obey to the eigenvalue equation $\left( 14\right) .$ \\

A simple computation leads to
\begin{equation}
\left| z,\alpha \right\rangle ={\cal N(}\left| z\right| {\cal )}%
\sum\limits_{n=0}^\infty \sqrt{\frac{\Gamma \left( 2+\frac
2{3\varepsilon }\right) }{\left( 3\varepsilon \right) ^n\Gamma
\left( n+1\right) \Gamma \left( n+2+\frac 2{3\varepsilon }\right)
}}\left( z\sqrt{2}\right) ^ne^{-i\alpha e_n}\left| n,\varepsilon
\right\rangle
\end{equation}
where
\begin{equation}
{\cal N(}\left| z\right| {\cal )}=\left[ \hbox{ }_0F_1\left(
2+\frac 2{3\varepsilon },\frac 2{3\varepsilon }\left| z\right|
^2\right) \right] ^{-\frac 12}.
\end{equation}
We remark that the coherent states $\left| z,\alpha \right\rangle
$ are continuously labeled by $z$ and $\alpha $, and the radius of
convergence is infinite. The measure in respect which we have an
overcomplet set of coherent states is
\begin{equation}
d\mu \left( z\right) =\frac 4{3\pi \varepsilon }I_{\left( 1+\frac
2{3\varepsilon }\right) }\left( 2\sqrt{\frac 2{3\varepsilon
}}r\right) K_{\left( \frac 12+\frac 1{3\varepsilon }\right)
}\left( 2\sqrt{\frac 2{3\varepsilon }}r\right) rdrd\phi,
\hspace{1cm}z=re^{i\phi }.
\end{equation}
The overlapping between two $x^4-$anharmonic oscillator coherent states is
given by

\begin{equation}
\left\langle z,\alpha \right. \left| z^{\prime },\alpha \right\rangle =\frac{%
_0F_1\left( 2+\frac 2{3\varepsilon },\frac 2{3\varepsilon }\overline{z}%
z^{\prime }\right) }{\sqrt{_0F_1\left( \frac 2{3\varepsilon }+2,\frac
2{3\varepsilon }\left| z\right| ^2\right) _0F_1\left( \frac 2{3\varepsilon
}+2,\frac 2{3\varepsilon }\left| z^{\prime }\right| ^2\right) }}.
\end{equation}
The Gazeau-Klauder coherent states provide a representation of any state \\$%
\left| \phi \right\rangle =\sum\limits_{n=0}^{+\infty }d_n\left|
n,\varepsilon \right\rangle $ in the Hilbert space by an entire
function

\begin{eqnarray}
\phi \left( z,\alpha \right) &=&\sqrt{_0F_1\left( 2+\frac 2{3\varepsilon
},\frac 2{3\varepsilon }\left| z\right| ^2\right) }\langle \overline{z}%
,\alpha \left| \phi \right\rangle  \nonumber \\
&=&\sum\limits_{n=0}^{+\infty }\sqrt{\frac{\Gamma \left( 2+\frac
2{3\varepsilon }\right) }{\left( 3\varepsilon \right) ^n\Gamma
\left(
n+1\right) \Gamma \left( n+2+\frac 2{3\varepsilon }\right) }}\left( z\sqrt{2}%
\right) ^nd_ne^{i\alpha e_n}.
\end{eqnarray}
The state $\left| n,\varepsilon \right\rangle $ is represented by

\begin{eqnarray}
\phi _{(n,\varepsilon )}\left( z,\alpha \right)
&=&\sqrt{_0F_1\left( 2+\frac 2{3\varepsilon },\frac 2{3\varepsilon
}\left| z\right| ^2\right) }\langle \overline{z},\alpha \left|
n,\varepsilon \right\rangle  \nonumber \\ &=&\sqrt{\frac{\Gamma
\left( 2+\frac 2{3\varepsilon }\right) }{\left( 3\varepsilon
\right) ^n\Gamma \left( n+1\right) \Gamma \left( n+2+\frac
2{3\varepsilon }\right) }}\left( z\sqrt{2}\right) ^ne^{i\alpha
e_n}.
\end{eqnarray}
The operators $A_\varepsilon ^{\pm }$ and $G_\varepsilon (N)=\left[
A_\varepsilon ^{-},A_\varepsilon ^{+}\right] =3\varepsilon (N+1)+1$ act in
the Hilbert space of analytic functions $\phi \left( z,\alpha \right) $ as
linear differential operators

\begin{equation}
A_\varepsilon ^{+}=z,\hspace{0.5cm}A_\varepsilon ^{-}=\frac{3\varepsilon }2z%
\frac{d^2}{dz^2}+\left( 1+3\varepsilon \right) \frac d{dz}\hspace{0.5cm}%
\hbox{and}\hspace{0.5cm}G_\varepsilon \left( N\right)
=3\varepsilon z\frac d{dz}+(1+3\varepsilon )
\end{equation}
with the following actions

\begin{eqnarray}
A_\varepsilon ^{+}\phi _{(n,\varepsilon )}\left( z,\alpha \right) &=&\sqrt{%
\left( \frac{3\varepsilon }2\right) \left( n+1\right) \left(
n+2+\frac 2{3\varepsilon }\right) }e^{-i\alpha (e_{n+1}-e_n)}\phi
_{(n+1,\varepsilon )}\left( z,\alpha \right),  \nonumber \\ && \\
A_\varepsilon ^{-}\phi _{(n,\varepsilon )}\left( z,\alpha \right) &=&\sqrt{%
\left( \frac{3\varepsilon }2\right) n\left( n+1+\frac
2{3\varepsilon }\right) }e^{i\alpha (e_n-e_{n-1})}\phi
_{(n-1,\varepsilon )}\left( z,\alpha \right), \\ G_\varepsilon
(N)\phi _{(n,\varepsilon )}\left( z,\alpha \right) &=&\left(
1+3\varepsilon \left( n+1\right) \right) \phi _{(n,\varepsilon
)}\left( z,\alpha \right).
\end{eqnarray}
Using the relation $\left( 25\right) ,$ we can write

\begin{equation}
\left| z,\alpha \right\rangle =\exp \left( z\frac 2{3\varepsilon
(N+1)+2}A_\varepsilon ^{+}\right) \left| 0,\varepsilon \right\rangle .
\end{equation}
Note that when $\varepsilon \rightarrow 0$, it is obvious that $x^4-$%
anharmonic oscillator leads to the harmonic oscillator Hamiltonian. Indeed,
using the formula

\begin{equation}
\lim_{\varepsilon \rightarrow 0}\left( \frac{3\varepsilon }2\right) ^n\frac{%
\Gamma (n+2+\frac 2{3\varepsilon })}{\Gamma (2+\frac 2{3\varepsilon })}=1,
\end{equation}
the Gazeau-Klauder coherent states Eq. $\left( 115\right) $ becomes

\begin{equation}
\left| z,\alpha \right\rangle =\exp \left( -\frac{\left| z\right| ^2}%
2\right) \sum\limits_{n=0}^{+\infty
}\frac{z^n}{\sqrt{n!}}e^{-i\alpha n}\left| n,0\right\rangle .
\end{equation}
We remark also that for the special case when $\varepsilon =\frac 23$ we get
all the result obtained for the infinite square well potential.

\subsection{Klauder-Perelomov coherent states}

Following the definition ${\bf D}_2$ the coherent states of
Klauder-Perelomov type for the nonlinear oscillators are defined as

\begin{equation}
\left| z,\alpha \right\rangle =\exp (zA_\varepsilon ^{+}-\overline{z}%
A_\varepsilon ^{-})\left| 0,\varepsilon \right\rangle,%
\hspace{0.5cm}z\in {\bf C.}
\end{equation}
After a more a less complicated manipulation or by applying the results of
the first section 1 one have

\begin{equation}
\left| z,\alpha \right\rangle =\sum\limits_{n=0}^{+\infty
}Z^nc_n\left( \left| Z\right| \right) \sqrt{F(n)}e^{-i\alpha
e_n}\left| n,\varepsilon \right\rangle
\end{equation}
where
\begin{equation}
Z=\sqrt{\frac{3\varepsilon }2}z\hspace{1cm}\hbox{and }\hspace{1cm}F(n)=\frac{%
\Gamma (n+1)\Gamma (n+2+\frac 2{3\varepsilon })}{\Gamma (2+\frac
2{3\varepsilon })}.
\end{equation}
The quantities $c_n\left( \left| Z\right| \right) $ are defined by
\begin{equation}
c_n\left( \left| Z\right| \right) =\sum\limits_{j=0}^{+\infty
}\frac{\left( -\left| Z\right| ^2\right) ^j}{\left( n+2j\right)
!}\pi (n+1,j)
\end{equation}
where
\begin{equation}
\pi (n+1,j)=\sum\limits_{i_1=1}^{n+1}e_{i_1}^{\prime
}\sum\limits_{i_2=1}^{i_1+1}e_{i_2}^{\prime
}....\sum\limits_{i_j=1}^{i_{j-1}+1}e_{i_j}^{\prime },\hbox{%
and }\hspace{0.3cm}\pi \left( n+1,0\right) =1
\end{equation}
with
\begin{equation}
e_n^{\prime }=\frac 2{3\varepsilon }e_n.
\end{equation}
We can verify that the $\pi $'s satisfy the following relation
\begin{equation}
\pi \left( n+1,j\right) -\pi \left( n,j\right) =\left( n+1\right) \left(
n+2+\frac 2{3\varepsilon }\right) \pi \left( n+2,j-1\right) .
\end{equation}
Note that here we have the same relations that ones obtained in
section 1 (see $\left( 30\right) $) with a minor modifications.
Using the recurrence
formula $\left( 134\right) $, it is not difficult to show that the $%
c_n\left( \left| Z\right| \right) ^{\prime }$s satisfy the following
differential equation
\begin{equation}
\left| Z\right| \frac{dc_n\left( \left| Z\right| \right) }{dr}=c_{n-1}\left(
\left| Z\right| \right) -nc_n\left( \left| Z\right| \right) -\left(
n+1\right) \left( n+2+\frac 2{3\varepsilon }\right) \left| Z\right|
^2c_{n+1}\left( \left| Z\right| \right) .
\end{equation}
Setting

\begin{equation}
c_n\left( \left| Z\right| \right) =\frac 1{n!\left| Z\right| ^n}{\cal A}%
_n\left( \left| Z\right| \right)
\end{equation}
the differential equation $\left( 135\right) $ takes the simple form
\begin{equation}
\frac{d{\cal A}_n\left( \left| Z\right| \right) }{d\left| Z\right| }=n{\cal A%
}_{n-1}\left( \left| Z\right| \right) -\left( n+2+\frac 2{3\varepsilon
}\right) {\cal A}_{n+1}\left( \left| Z\right| \right) .
\end{equation}
It follows that the solutions of $\left( 137\right) $ are

\begin{equation}
{\cal A}_n\left( \left| Z\right| \right) =\left[ \cosh (\left|
Z\right| )\right] ^{-n-2-\frac 2{3\varepsilon }}\left[ \sinh
(\left| Z\right| )\right] ^n.
\end{equation}
Finally the coherent states Eq. $\left( 128\right) $ are given by

\begin{equation}
\left| \zeta ,\alpha \right\rangle \equiv \left( 1-\left| \zeta \right|
^2\right) ^{1+\frac 1{3\varepsilon }}\sum\limits_{n=0}^{+\infty }\frac{%
\zeta ^n}{\sqrt{n!}}\left[ \frac{\Gamma \left( n+2+\frac 2{3\varepsilon
}\right) }{\Gamma (2+\frac 2{3\varepsilon })}\right] ^{\frac 12}e^{-i\alpha
e_n}\left| n,\varepsilon \right\rangle
\end{equation}
where $\zeta =\frac{Z\tanh \left( \left| Z\right| \right) }{\left| Z\right| }%
,$ and they form an overcomplet set in respect to measure
\begin{equation}
d\mu \left( \zeta \right) =\left( \frac 1\pi +\frac 2{3\varepsilon \pi
}\right) \frac{d^2\zeta }{\left( 1-\left| \zeta \right| ^2\right) ^2}.
\end{equation}
The kernel $\left\langle \zeta ,\alpha \right| \zeta ^{^{\prime }},\alpha
^{^{\prime }}\rangle $ is easily evaluated from $\left( 139\right) $%
\begin{eqnarray}
\left\langle \zeta ,\alpha \right| \zeta ^{^{\prime }},\alpha
^{^{\prime }}\rangle &=&\sqrt{\left( 1-\left| \zeta \right|
^2\right) ^{1+\frac 1{3\varepsilon }}\left( 1-\left| \zeta
^{^{\prime }}\right| ^2\right) ^{1+\frac 1{3\varepsilon
}}}\sum\limits_{n=0}^{+\infty }\frac{\left( \overline{\zeta }\zeta
^{^{\prime }}\right) ^n}{n!}\times  \nonumber \\ &&\left[
\frac{\Gamma \left( n+2+\frac 2{3\varepsilon }\right) }{\Gamma
(2+\frac 2{3\varepsilon })}\right] e^{-i(\alpha ^{^{\prime
}}-\alpha )n(n+2)}.
\end{eqnarray}
For an arbitrary state $\left| \varphi \right\rangle
=\sum\limits_{n=0}^{+\infty }f_n\left| n,\varepsilon \right\rangle
\in {\cal H}$, one can construct the analytic function

\begin{equation}
\varphi (\zeta {\bf ,}\alpha )=\left( 1-\left| \zeta \right|
^2\right) ^{-1-\frac 1{3\varepsilon }}\left\langle \overline{\zeta
},\alpha \right| \varphi \rangle =\sum\limits_{n=0}^{+\infty
}\zeta ^n\left[ \frac{\Gamma
\left( n+2+\frac 2{3\varepsilon }\right) }{\Gamma (2+\frac 2{3\varepsilon })}%
\right] ^{\frac 12}f_ne^{i\alpha e_n}
\end{equation}
with
\begin{equation}
\left| \varphi \right\rangle =\int\left| \zeta ,\alpha
\right\rangle
\left( 1-\left| \zeta \right| ^2\right) ^{1+\frac 1{3\varepsilon }}\varphi (%
\overline{\zeta }{\bf ,}\alpha )d\mu \left( \zeta \right) .
\end{equation}
In particular, for the states $\left| n,\varepsilon \right\rangle $ we
associate the monomial

\begin{equation}
\phi _{(n,\varepsilon )}^{\prime }(\zeta ,\alpha )=\zeta ^n\left[ \frac{%
\Gamma \left( n+2+\frac 2{3\varepsilon }\right) }{\Gamma (2+\frac
2{3\varepsilon })}\right] ^{\frac 12}e^{i\alpha e_n}.
\end{equation}
The creation $A_\varepsilon ^{+}$ annihilation $A_\varepsilon ^{-}$ and $%
G_\varepsilon (N)$ operators act on the Hilbert space of analytic functions $%
\varphi (\zeta {\bf ,}\alpha )$ as follows

\begin{equation}
A_\varepsilon ^{+}=\sqrt{\frac{3\varepsilon }2}\left[ \zeta ^2\frac d{d\zeta
}+(2+\frac 2{3\varepsilon })\zeta \right],\hspace{0.8cm}%
A_\varepsilon ^{-}=\sqrt{\frac{3\varepsilon }2}\frac d{d\zeta }
\end{equation}
and
\begin{equation}
G_\varepsilon (N)=3\varepsilon \left[ \zeta \frac d{d\zeta }+1+\frac
1{3\varepsilon }\right] .
\end{equation}
One can verify that

\begin{eqnarray}
A_\varepsilon ^{+}\phi _{(n,\varepsilon )}^{\prime }(\zeta ,\alpha ) &=&%
\sqrt{\left( \frac{3\varepsilon }2\right) \left( n+1\right) \left(
n+2+\frac 2{3\varepsilon }\right) }e^{-i\alpha (e_{n+1}-e_n)}\phi
_{(n+1,\varepsilon )}^{\prime }(\zeta ,\alpha )  \nonumber, \\ &&
\\
A_\varepsilon ^{-}\phi _{(n,\varepsilon )}^{\prime }(\zeta ,\alpha ) &=&%
\sqrt{\left( \frac{3\varepsilon }2\right) n\left( n+1+\frac
2{3\varepsilon }\right) }e^{i\alpha (e_n-e_{n-1})}\phi
_{(n-1,\varepsilon )}^{\prime }(\zeta ,\alpha ), \\ G_\varepsilon
\left( N\right) \phi _{(n,\varepsilon )}^{\prime }(\zeta ,\alpha )
&=&\left( 1+3\varepsilon \left( n+1\right) \right) \phi
_{(n,\varepsilon )}^{\prime }(\zeta ,\alpha ).
\end{eqnarray}
Now, let us discuss the limit $\varepsilon \rightarrow 0.$

By using the formula $\left( 126\right) ,$ the coherent states Eq. $\left(
139\right) $ takes the form

\begin{equation}
\left| z,\alpha \right\rangle =\exp \left( -\frac{\left| z\right| ^2}%
2\right) \sum\limits_{n=0}^{+\infty
}\frac{z^n}{\sqrt{n!}}e^{-i\alpha n}\left| n,0\right\rangle
\end{equation}
and the measure, Eq. $\left( 140\right) ,$ becomes

\begin{equation}
\lim_{\varepsilon \rightarrow 0}d\mu \left( \zeta \right) =d\mu \left(
z\right) =\frac 1\pi d^2z,
\end{equation}
we note also that for $\varepsilon \rightarrow 0$ the representation of the
operators $A_\varepsilon ^{+}$ and $A_\varepsilon ^{-}$ reduces to creation
and annihilation ones of the harmonic oscillator (Weyl-Heisenberg algebra).
Indeed, we have

\begin{equation}
A_\varepsilon ^{+}\rightarrow a^{+}\equiv z\hspace{0.2cm},\hspace{0.4cm}%
A_\varepsilon ^{-}\rightarrow a^{-}\equiv \frac d{dz}\hspace{0.4cm}\hbox{and}%
\hspace{0.3cm}G_\varepsilon (N)\rightarrow I
\end{equation}

\subsection{Generalized intelligent states}

Having the analytical representation of the Gazeau-Klauder and
Klauder-Perelomov coherent states, we can derive the generalized
intelligent states, for the $x^4-$anharmonic oscillator, in an
analytical ways.

\subsubsection{Gazeau-Klauder analytic representation}

The Hilbert space, in this representation, is the space of analytic function
\{$\phi \left( z,\alpha \right) $\} equipped with the scalar product Eq. $%
\left( 83\right) ,$ where the measure $d\mu \left( z\right) $ is given by $%
\left( 117\right) .$

To solve the eigenvalue equation $\left( 38\right) $ in the case of $x^4-$%
anharmonic oscillators, we introduce the function

\begin{equation}
\phi _{(z^{\prime },\lambda ,\alpha )}\left( z,\alpha \right) =\sqrt{%
_0F_1\left( 2+\frac 2{3\varepsilon },\frac 2{3\varepsilon }\left| z\right|
^2\right) }\langle \overline{z},\alpha \left| z^{\prime },\lambda ,\alpha
\right\rangle .
\end{equation}
Then the eigenvalue equation $\left( 38\right) ,$ is converted to the
second-order homogenous differential equation

\begin{equation}
\left\{ \left( \left( 1+\lambda \right) \frac{3\varepsilon }2\right) \left[
\frac 2{3\varepsilon }\left( 1+3\varepsilon \right) \frac d{dz}+z\frac{d^2}{%
dz^2}\right] +\left( 1-\lambda \right) z\right\} \phi _{\left( z^{\prime
},\lambda \right) }(z)=2z^{\prime }\phi _{\left( z^{\prime },\lambda \right)
}(z).
\end{equation}
By means of simple substitutions, the above equation is reduced to the
Kummer equation for the confluent hypergeometric function $_1F_1\left(
a;b;z\right) $ $\left[ 19\right] $ and we arrive to the following solution
\begin{equation}
\phi _{\left( z^{\prime },\lambda \right) }(z)=\exp \left( cz\right) \hbox{ }%
_1F_1\left( a;b;-2cz\right)
\end{equation}
where
\begin{equation}
a=1+\frac 1{3\varepsilon }\mp \frac{2z^{\prime }}{\sqrt{\left(
\lambda ^2-1\right) 6\varepsilon }}\hbox{,}\hspace{0.5cm}b=\frac
2{3\varepsilon }+2\hspace{0.3cm}\hbox{and }\hspace{0.4cm}c=\pm
\sqrt{\left( \frac{\lambda -1}{\lambda +1}\right) \frac
2{3\varepsilon }}.
\end{equation}
We note that the generalized intelligent states for the harmonic oscillator
can be obtained from the equation $\left( 155\right) $ in the limit $%
\varepsilon \rightarrow 0$ (or from the differential equation $\left(
154\right) $ by setting $\varepsilon =0$). Thus, we have
\begin{equation}
\phi _{\left( z^{\prime },\lambda \right) }(z)=\phi _{\left( z^{\prime
},\lambda \right) }(0)\exp \left( \frac{2z^{\prime }}{1+\lambda }z+\left(
\frac{\lambda -1}{\lambda +1}\right) \frac{z^2}2\right)
\end{equation}
where $\phi _{\left( z^{\prime },\lambda \right) }(0)$ is the normalization
constant.

\subsubsection{Klauder-Perelomov analytic representation}

In this representation, the Hilbert space is equipped with the scalar
product given by $\left( 94\right) $ and the measure $d\mu (\zeta )$ is
given by $\left( 140\right) .$

By introducing the analytic function
\begin{equation}
\varphi _{(\zeta ^{\prime },\lambda )}\left( \zeta \right) =\left( 1-\left|
\zeta \right| ^2\right) ^{-1-\frac 1{3\varepsilon }}\langle \overline{\zeta }%
,\alpha \left| \zeta ^{\prime },\lambda ,\alpha \right\rangle
\end{equation}
we convert the eigenvalue equation $\left( 38\right) $ into the differential
equation

\begin{equation}
\left[ \left[ (1-\lambda )\zeta ^2+(1+\lambda )\right] \frac d{d\zeta
}+(1-\lambda )\left( 2+\frac 2{3\varepsilon }\right) \zeta -2\sqrt{\frac
2{3\varepsilon }}\zeta ^{\prime }\right] \varphi _{(\zeta ^{\prime },\lambda
)}(\zeta )=0.
\end{equation}
In the general case where $\lambda \neq \pm 1,$ the solution of the equation
$\left( 159\right) $ is

\begin{equation}
\varphi _{(\zeta ^{\prime },\lambda )}(\zeta )={\cal B}\left( \left| \zeta
\right| \right) \left( 1+\left( \frac{\lambda -1}{\lambda +1}\right) ^{\frac
12}\zeta \right) ^{\alpha _{+}}\left( 1-\left( \frac{\lambda -1}{\lambda +1}%
\right) ^{\frac 12}\zeta \right) ^{\alpha _{-}}
\end{equation}
where
\begin{equation}
\alpha _{\pm }=-1-\frac 1{3\varepsilon }\pm \frac{2\zeta ^{\prime }}{\sqrt{%
\left( \lambda ^2-1\right) 6\varepsilon }}
\end{equation}
and ${\cal B}\left( \left| \zeta \right| \right) $ is a normalization
constant.

The condition of the analyticity of the solution $\varphi _{(\zeta
^{\prime },\lambda )}(\zeta )$ in the unit disk is satisfied when
we have

\begin{equation}
\left| \frac{\lambda -1}{\lambda +1}\right| <1\Leftrightarrow
\hbox{ }Re\lambda >0.
\end{equation}

In order to obtain the decomposition of the generalized intelligent states $%
\left| \zeta ^{\prime },\lambda ,\alpha \right\rangle $ over the Hilbert
orthonormal basis \{$\left| n,\varepsilon \right\rangle $\}, we expand the
function $\varphi _{(\zeta ^{\prime },\alpha )}(\zeta )$ into a power series
in $\zeta $ in the same way that discussed previously for the infinite
square well potential. This can be done by using the following relations

\begin{equation}
\prod\limits_{l=\pm 1}\left( 1+\left( \frac{\lambda -1}{\lambda
+1}\right)
^{\frac 12}\zeta \right) ^{-1-\frac 1{3\varepsilon }+l\frac{2\zeta ^{\prime }%
}{\sqrt{(\lambda ^2-1)6\varepsilon }}}=\sum\limits_{n=0}^{+\infty
}\zeta ^n\left( 2\sqrt{\frac{\lambda -1}{\lambda +1}}\right)
^nP_n^{(\alpha _{+}-n,\alpha _{-}-n)}(0)
\end{equation}
where $P_n^{(\alpha ,\beta )}(x)$ is the Jacobi polynomials $\left[
19\right] $.

Using the relation between the hypergeometric function and the Jacobi
polynomials $\left[ 19\right] $, one can show that

\begin{eqnarray}
\left| \zeta ^{\prime },\lambda ,\alpha \right\rangle &=&{\cal B}\left(
\left| \zeta \right| \right) \sum\limits_{n=0}^{+\infty }\left[ \frac{n!}{%
(n+2)!}\right] ^{\frac 12}\left[ \frac{n!\Gamma (\alpha _{+}-n+1)}{\Gamma
(\alpha _{+}+1)}\right] \times \\
&&\left( 2\sqrt{\frac{\lambda -1}{\lambda +1}}\right) ^n\hbox{ }%
_2F_1(-n,-n-2;\alpha _{+}-n+1;\frac 12)e^{-i\alpha e_n}\left| \psi
_n\right\rangle .  \nonumber
\end{eqnarray}
The two generalized intelligent states $\phi _{\left( z^{\prime
},\lambda \right) }(z)$ (Eq. $\left( 155\right) $) and $\varphi
_{(\zeta ^{\prime },\lambda )}(\zeta )$ (Eq. $\left( 160\right) $)
are related as in the previous case (i. e.,the infinite square
well potential) through the Laplace transform.

\section{Concluding remarks}

In this paper, we construced the coherent states (\`{a} la Gazeau-Klauder
and \`{a} la Klauder-Perelomov) and the generalized intelligent states for
an arbitrary quantum systems. As an illustration of our construction, we
treated the system of a free particle in the infinite square well potential
and the $x^4-$nonlinear oscillators. We shown the advantage of the analytic
representations of Gazeau-Klauder as well as Klauder-Perelomov coherent
states in obtaining the generalized intelligent states in analytical ways.
Finally, one can see that our results could be extended to other exactly
solvable quantum systems like, for instance, coulomb, hyperbolic
Rosen-Morse, Eckart and trigonometric Rosen-Morse potentials. This extension
is under study.\vspace{1.5cm}

{\bf Acknowledgements}

The authors would like to thank the Abdus Salam ICTP (Trieste-Italy) for
their hospitality and M. Daoud is grateful to Professor Yu Lu for his kind
invitation to joint the condenced matter section of the AS-ICTP.

\newpage\

\end{document}